\documentclass[aps,prl,twocolumn,longbibliography,superscriptaddress,notitlepage,floatfix]{revtex4-2}
\renewcommand{\figurename}{Fig.}
\usepackage{amsmath,amsfonts,amssymb,color,epsfig,graphics,graphicx,latexsym,theorem,url,multirow,diagbox}
\usepackage[colorlinks,colorlinks,citecolor=blue,linkcolor=blue,urlcolor=blue]{hyperref}
\usepackage[utf8]{inputenc}
\usepackage{booktabs}
\usepackage[T1]{fontenc}
\usepackage{courier}
\usepackage{listings}
\usepackage{latexsym}
\usepackage{dcolumn}
\usepackage{comment}
\usepackage{times} 
\usepackage[ruled,vlined,linesnumbered]{algorithm2e}
\usepackage{braket}
\usepackage{bm}
\usepackage{siunitx}

\definecolor{blue}{rgb}{0,0,1}
\definecolor{red}{rgb}{1,0,0}
\definecolor{green}{rgb}{0,1,0}
\definecolor{sec_red}{RGB}{166, 40, 37}

\begin{document}

\title{
Machine learning the arrow of time in solid-state spins
}

\affiliation{
Center for Quantum Information, IIIS, Tsinghua University, Beijing 100084, China\\
$^{2}$ Shanghai Qi Zhi Institute, Shanghai 200232, China\\
$^{3}$ Hefei National Laboratory, Hefei 230088, China\\
$^{4}$ International Research Centre MagTop, Institute of Physics, Polish Academy of Sciences, Aleja Lotnikow 32/46, PL-02668 Warsaw, Poland} 

\author{Xiang-Qian Meng$^{1}$}\thanks{These authors contributed equally to this work.}
\author{Zhide Lu$^{2}$}\thanks{These authors contributed equally to this work.}
\author{Ya-Nan Lu$^{1,3}$}
\author{Xiu-Ying Chang$^{1,3}$}
\author{Yan-Qing Liu$^{1,3}$}
\author{Dong Yuan$^{1}$}
\author{Weikang Li$^{1}$}
\author{Zheng-Zhi Sun$^{1}$}
\author{Pei-Xin Shen$^{4}$}
\author{Lu-Ming Duan$^{1,3}$}\email{lmduan@tsinghua.edu.cn}
\author{Dong-Ling Deng$^{1,2,3}$}\email{dldeng@tsinghua.edu.cn}
\author{Pan-Yu Hou$^{1,3}$}\email{houpanyu@tsinghua.edu.cn}

\begin{abstract}{
\noindent
Understanding the emergence of the thermodynamic arrow of time in microscopic systems is of fundamental importance, particularly given that unitary evolution preserves time-reversal symmetry.
While projective measurements introduce temporal irreversibility, identifying this asymmetry from single evolution trajectories in the presence of stochastic fluctuations presents a considerable challenge.
Here, we harness machine learning to identify the arrow of time from individual trajectories generated by a programmable ten-qubit quantum processor based on a nitrogen-vacancy center in diamond. 
We implement quantum circuits that realize unitary evolutions where heat flows from hotter to colder subsystems and their time-reversed counterparts.
Projective measurements inserted in these processes induce entropy production, and their outcomes constitute the evolution trajectory. 
We demonstrate that an unsupervised clustering algorithm autonomously divides the experimental trajectories into two distinct groups without prior knowledge, while a convolutional neural network identifies the temporal direction of these trajectories with approximately 92\% accuracy. 
In addition, we show that a diffusion-based generative model reproduces essential signatures of directional energy flow and entropy production. 
Our results establish machine learning as a powerful tool for uncovering underlying physical processes from complex experimental data, advancing the interface between quantum thermodynamics and artificial intelligence.
}
\end{abstract}

\maketitle

\noindent
The second law of thermodynamics, a cornerstone of physics, dictates that heat spontaneously flows from hot to cold systems and that entropy production is non-negative on average~\cite{landau1980statistical}.
This fundamental irreversibility underlies the ``arrow of time'' famously coined by Eddington~\cite{eddington2019nature}---a concept that captures the asymmetry of macroscopic processes.
In isolated quantum systems, dynamics governed by the system Hamiltonian are fully reversible and preserve von Neumann entropy. This presents an apparent paradox: how does macroscopic irreversibility emerge from microscopical reversibility?
Resolutions to this paradox often invoke the role of environment or the act of measurements~\cite{schulman1997time}. In particular, projective measurements break the time-reversal symmetry through entropy production~\cite{Nielsen2010Quantum}, establishing a preferred temporal direction (Fig.~\ref{fig:1}\textbf{a}).

However, identifying the arrow of time at the microscopic scale is non-trivial.
Unlike macroscopic systems, where irreversibility is deterministic and evident, microscopic systems are subject to stochastic fluctuations~\cite{Jarzynski1997Nonequilibrium,Crooks1999Entropy,Esposito2009Nonequilibrium,Campisi2011Colloquium,Landi2021Irreversible}, making ``reverse'' events (where entropy seemingly decreases) statistically possible, albeit rare.
To reliably distinguish the true temporal direction amidst these fluctuations requires analyzing complex, high-dimensional statistical distributions.
Recently, machine learning as a powerful tool for scientific research~\cite{Sarma2019Machine,Carleo2019Machine,Dawid2025Machine} has exhibited remarkable capabilities in recognizing patterns within high-dimensional data~\cite{Zhang2019Machine,Wu2024Universal}.
For instance, it has been employed to assist studies in quantum physics, including the classification of phases of matter~\cite{Carrasquilla2017Machine,VanNieuwenburg2017Learning}, quantum state tomography~\cite{Torlai2018Neuralnetwork}, and the optimization of experimental controls~\cite{Niu2019Universal,Sivak2022ModelFree}.
Given its ability to extract subtle features, machine learning offers a feasible avenue to identify the arrow of time from microscopic data, as recently explored in theoretical studies~\cite{Seif2021Machine}.
However, it remains unclear whether this methodology can be extended to quantum systems, particularly when applied to experimental data in the presence of imperfections and noise.
Here, we demonstrate that machine learning models autonomously identify the arrow of time hidden in the experimental trajectories of quantum thermodynamic processes without prior knowledge.

\begin{figure*}[t]
\includegraphics[width=0.8\textwidth]{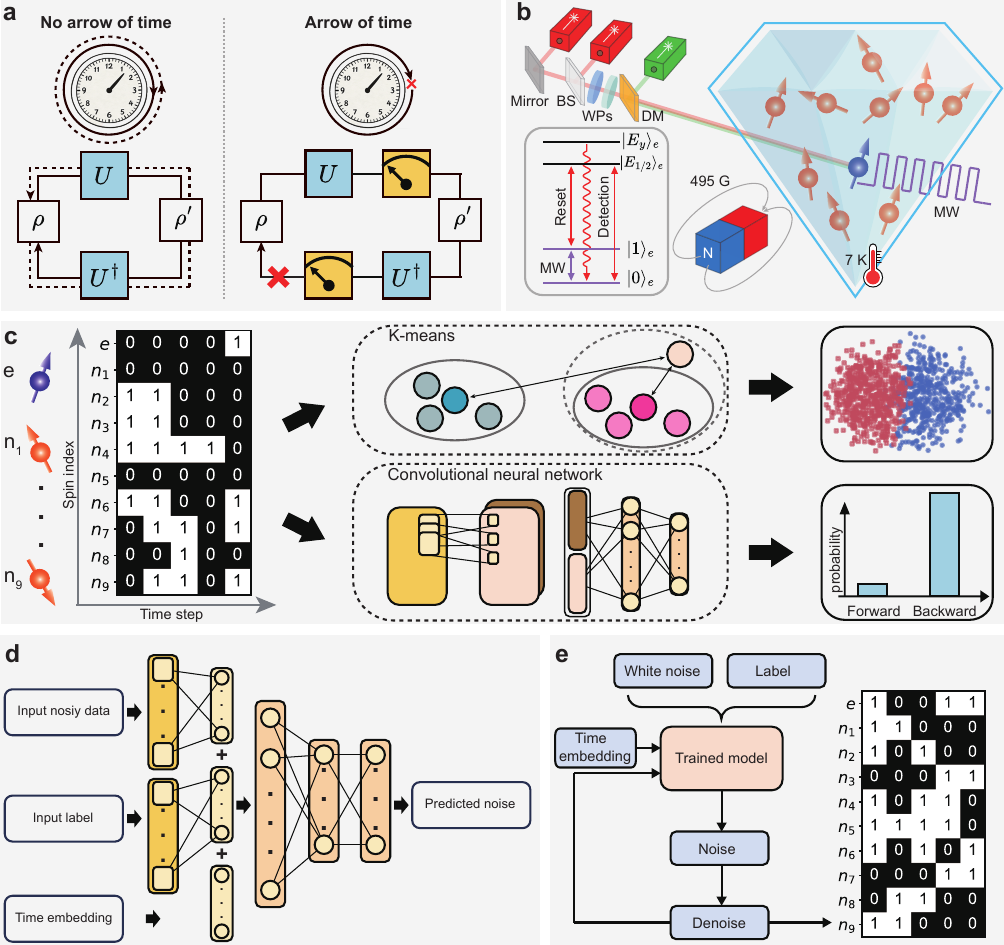}
\caption{
\textbf{Machine learning of the time arrow in the evolution of quantum systems.}
\textbf{a}, Illustration of time irreversibility induced by projective measurements. Unitary evolution (blue boxes) preserves time-reversal symmetry, whereas the introduction of projective measurements (orange boxes) breaks this symmetry, establishing the arrow of time.
\textbf{b}, Schematic of the experimental system. A nitrogen-vacancy center system in diamond consists of a single electron spin and nine nuclear spins of nearby ${}^{13}\text{C}$ atoms. The system operates at approximately $7$~K and in an external magnetic field of approximately $495$~Gauss. A green laser at 532\,nm prepares the charge state of the nitrogen-vacancy center. Two 637\,nm lasers resonantly drive the transitions shown in the inset for electron spin initialization and readout. Microwave (MW) fields are tailored for coherent rotations of the electron spin and universal control of individual nuclear spins. The laser beams are directed onto the diamond via an optical setup comprising mirrors, beam splitters (BS), wave plates (WPs), and dichroic mirrors (DM) (Supplementary Information).
\textbf{c}, Machine learning frameworks for identifying the arrow of time. The input consists of a single evolution trajectory of the ten qubits over five sequential measurements, represented as a $[10 \times 5]$ binary matrix. This data is processed via two distinct pipelines: an unsupervised $k$-means clustering algorithm (upper middle) that autonomously groups trajectories into forward and backward clusters (upper right), and a supervised convolutional neural network (lower middle) that takes the matrix as input and outputs the classification probability for the temporal direction (lower right).
\textbf{d}, Architecture of the diffusion model for trajectory generation. The model consists of fully connected layers with sigmoid linear unit activation functions. Inputs include noisy data, class labels (forward or backward), and time embeddings. The network is trained to predict the injected noise.
\textbf{e}, Trajectory generation pipeline using the diffusion model. The model synthesizes trajectories consistent with physical data through a stepwise denoising process starting from the Gaussian white noise.
}
\label{fig:1}
\end{figure*}

\begin{figure*}[t]
\includegraphics[width=0.8\textwidth]{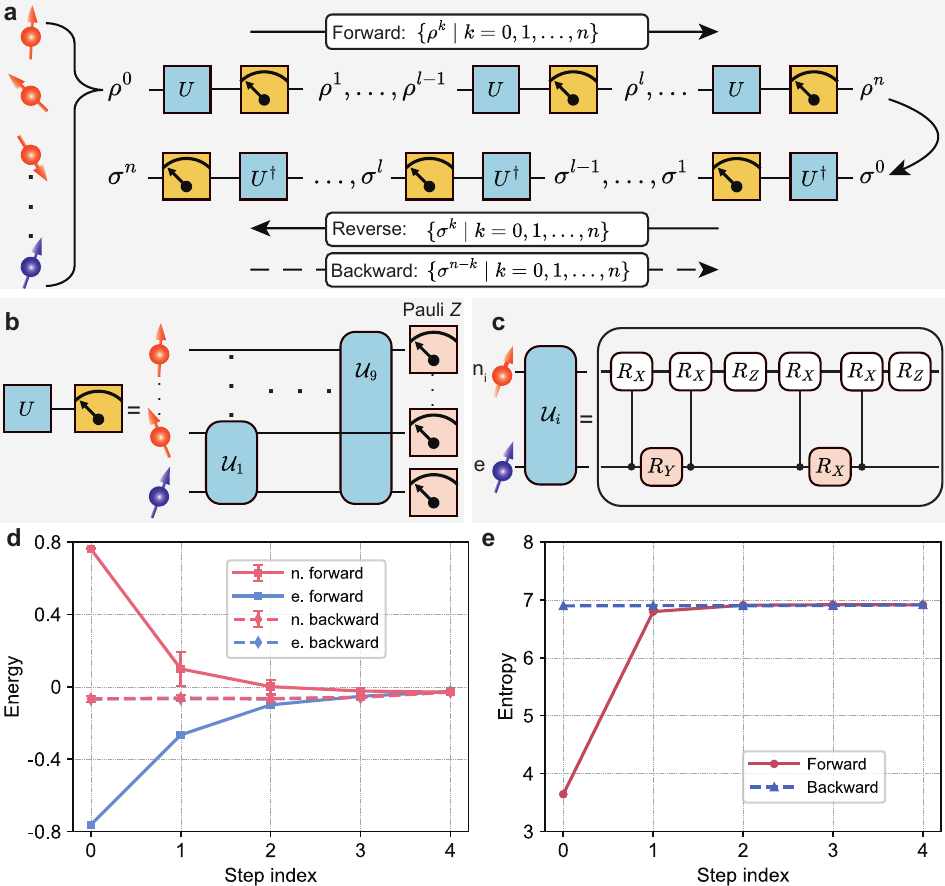}
\caption{\textbf{Experimental observation of the arrow of time in quantum thermodynamics.}
\textbf{a}, Schematic of quantum thermodynamic processes in a ten-spin system. In the forward process, spins are initialized in thermal Gibbs states at different temperatures. They then undergo repeated cycles of unitary evolution \textit{U} followed by projective measurements, yielding the forward state trajectory $\{\rho^{k}\}_{k=0}^{n}$.
The reverse process starts with the final state of the forward process. The system evolves similarly under unitary $U^\dagger$ in each step. The backward trajectory (dashed arrow) is obtained by time-reversing the trajectory of the reverse process. Due to the back-action of the measurements in our setup, the spins are re-prepared after each measurement according to the outcome.
\textbf{b}, Quantum circuit of unitary evolution. The unitary  $U$ is implemented as a sequence of pairwise unitaries $\mathcal{U}_i$ between the electron spin and the $i$th nuclear spin, followed by Pauli-$Z$ measurements.
\textbf{c}, The pairwise unitary $\mathcal{U}_i = \exp\left[-\mathrm{i}\left(X_\text{e} X_i + Y_\text{e} Y_i\right)\Delta t\right]$  with $\hbar=1$ comprises of single- and two-qubit gates. White blocks represent electron spin rotations by an angle of $\pi/2$, and pink blocks denote nuclear spin rotations by an angle of $2\Delta t$.
\textbf{d}, Measured energy of the electron spin (blue symbols) and average energy of the nine nuclear spins (red symbols) as functions of the step index for both forward (squares) and backward (diamonds) processes. Error bars represent one standard deviation of the mean over the nine nuclear spins.
\textbf{e}, Plots of the system entropy versus step index. The discrepancy between the forward (red circles) and backward (blue triangles) trajectories highlights the irreversibility of the thermodynamic process.
}
\label{fig:2}
\end{figure*}

We acquire experimental trajectories on a programmable quantum processor based on a nitrogen-vacancy (NV) center platform~\cite{Gruber1997Scanning,Jelezko2004Observation1,Jelezko2004Observation2,Meijer2005Generation,Rabeau2006Implantation}. It is a single defect in diamond with broad applications in quantum computing~\cite{Childress2006Coherent,Dutt2007Quantum,Hanson2008Coherenta,Neumann2008Multipartite,Balasubramanian2009Ultralong,Cappellaro2009Coherence,Neumann2010Quantum}, quantum networks~\cite{Bernien2012TwoPhoton,Bernien2013Heralded,Pfaff2014Unconditional,Hensen2015Loopholefree,Reiserer2016Robust,Kalb2017Entanglement,Humphreys2018Deterministic,Hermans2022Qubit} and quantum sensing~\cite{Maze2008Nanoscale,Balasubramanian2008Nanoscale,Taylor2008High,Degen2017Quantum}.
We implement a quantum circuit in such a system consisting of ten spin qubits to emulate quantum thermodynamic processes where heat flows from hot qubits to cold qubits (Fig.~\ref{fig:1}\textbf{b}). We perform multiple projective measurements in the process and record the sequential outcomes, forming evolution trajectories, referred to as forward trajectories. 
We also perform a time-reversal evolution with projective measurements and record their outcomes as backward trajectories.
We use machine learning algorithms to learn the arrow of time in these experimental trajectories. 
In particular, we demonstrate that an unsupervised clustering algorithm (Fig.~\ref{fig:1}\textbf{c}) autonomously distinguishes forward from backward trajectories with an accuracy exceeding $90\%$. 
We also employ a supervised convolutional neural network (Fig.~\ref{fig:1}\textbf{c}) to identify the arrow of time in those trajectories with an accuracy of approximately $92\%$. 
In addition, we demonstrate that a diffusion-based generative model (Fig.~\ref{fig:1}\textbf{d}, \textbf{e}) learns the underlying thermodynamic physics. After training on the experimental dataset, the model generates trajectories that reproduce irreversible heat flow and entropy production, consistent with experimental outcomes.
These results show that the machine learning models have captured the intrinsic temporal symmetry breaking that defines the arrow of time.

\begin{figure*}[t]
\includegraphics[width=\textwidth]{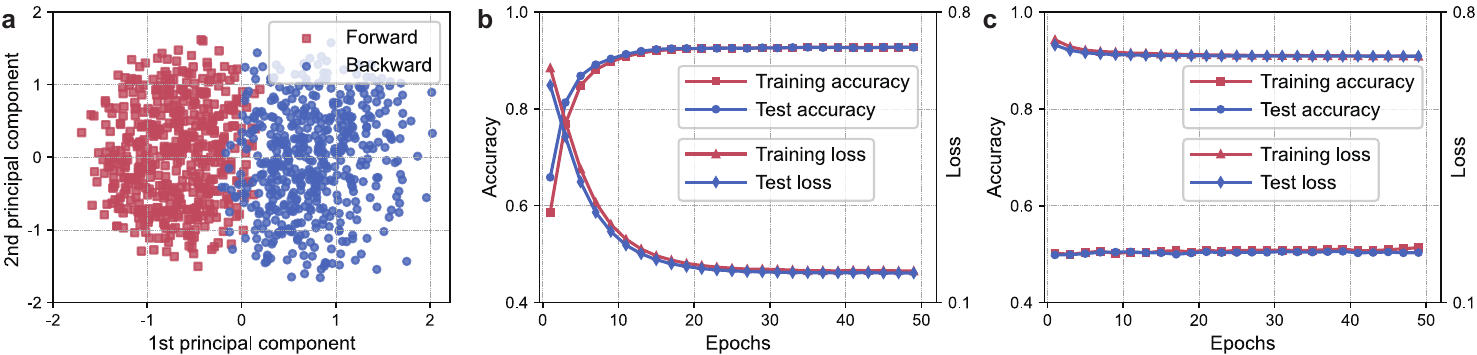}
\caption{
\textbf{Identification of the arrow of time by machine learning.}
\textbf{a}, An unsupervised clustering $k$-means algorithm without label supervision classifies $1,000$ experimental trajectories into two clusters.
These clusters correspond to the forward (red squares) and backward (blue circles) trajectories and are separated in the subspace spanned by the first two principal components.
\textbf{b}, Supervised learning on experimental data using a convolutional neural network distinguishes forward from backward trajectories with accuracy up to about 92\%. The loss function used here is the binary cross-entropy loss function.
\textbf{c}, Learning curves for simulated trajectories governed by pure unitary dynamics without projective measurements. The classification accuracy using the convolutional neural network collapses to the random-guessing baseline around $50\%$.
}
\label{fig:3}
\end{figure*}

\vspace{.6cm}
\noindent\textbf{\large{}Framework and experimental setup} \\
We first introduce the general framework for machine learning of the arrow of time. As shown in Fig.~\ref{fig:2}\textbf{a}, we consider two subsystems, a central qubit as subsystem A and a bath of $N$ qubits as subsystem B. The local Hamiltonian is defined as $H_A=Z_A$, $H_B=\sum_{i=1}^N Z_i$, where $Z$ denotes the Pauli-$Z$ operator of a single qubit.
Initially, both subsystems are prepared in independent thermal Gibbs states,
$\rho_{\text{AB}}^0 = \rho_{\text{A}}^0 \otimes \rho_{\text{B}}^0$, where $\rho_{\text{A}}^0 = e^{-\beta_{\text{A}} Z_{\text{A}}}/\mathcal{Z}_{\text{A}}$ and $\rho_{\text{B}}^0 = \bigotimes_{i=1}^N \left[e^{-\beta_i Z_i}/\mathcal{Z}_i\right]$ denote the initial states of subsystems A and B, characterized by inverse temperatures $\beta_\text{A}$ and $\beta_i$. $\mathcal{Z}_j = \operatorname{Tr}\left[\exp\left(-\beta_j Z_j\right)\right]$ ($j = \text{A}\ \text{or}\ j\in \left\{1,...,N\right\}$) represents the partition function of a single qubit.
As shown in Fig.~\ref{fig:2}\textbf{b}, the unitary evolution of the system is engineered as $U=\prod_{i=N}^1 \mathcal{U}_{i}$, where $\mathcal{U}_{i}=\exp\left(-\mathrm{i}H_i\Delta t\right)$ is generated by the exchange interaction $H_i= X_\text{e} X_{i} + Y_\text{e} Y_{i}$ between subsystem A and the $i$-th qubit of subsystem B. 
While unitary evolutions preserve von Neumann entropy and are therefore reversible in principle, the introduction of projective measurements breaks this time-reversal symmetry, establishing an arrow of time via entropy production. 
The considered thermodynamic protocol consists of a forward process and a reverse process, both including multiple steps. 
In each step of the forward process, subsystem A interacts with subsystem B via $U$, followed by projective measurements of each qubit in the Pauli-$Z$ basis. 
A salient feature of this process is the entropy increase: $S(\rho^{k}_{\text{AB}}) \geq S(\rho^{k-1}_{\text{AB}})$, where the von Neumann entropy is defined as $S(\rho) = -\operatorname{Tr}\left[\rho \ln \left(\rho\right)\right]$ and $k$ is the index of step. Repeating this for $n$ steps yields the forward state trajectory $\boldsymbol{\rho}^{F} = \left\{ \rho^{k}_{\mathrm{AB}} \right\}_{k=0}^n$.
The reverse process starts with the final state of the forward process, $\sigma^0_{\text{AB}} = \rho^{n}_{\text{AB}}$. In each step, the system evolves via the inverse operator $U^{\dagger}$ and each qubit is subsequently measured in the Pauli-$Z$ basis, yielding the reverse state trajectory $\boldsymbol{\sigma}^{R} = \left\{ \sigma^{k}_{\mathrm{AB}} \right\}_{k=0}^n$. We explicitly define the ``backward trajectory'' as the time-reversal of this sequence, denoted as $\overline{\boldsymbol{\sigma}^{R}} = \left\{ \sigma^{n-k}_{\mathrm{AB}} \right\}_{k=0}^n$. In this framework, the entropy increases in the forward state trajectory $\boldsymbol{\rho}^{F}$, in stark contrast, decreases in the backward state trajectory $\overline{\boldsymbol{\sigma}^{R}}$, representing the thermodynamic irreversibility introduced by projective measurements. 

From a thermodynamic perspective, the interaction operator commutes with the total Hamiltonian of the interacting pair, i.e., $\left[H_{i}, H_{\text{A}} + H_B\right] = 0$, ensuring that the unitary $\mathcal{U}_{i}$ performs no work on the total system~\cite{Micadei2019Reversing}. Consequently, the total energy is conserved, and any local energy variation is attributed to heat exchange.
In the framework mentioned above, heat flows naturally from hotter qubit to colder qubit in the presence of projective measurements, establishing the arrow of time.

As illustrated in Fig.~\ref{fig:1}\textbf{b}, our experiments are conducted with ten solid-state spins associated with a single NV center in diamond. 
The system operates under cryogenic conditions in an external magnetic field of approximately $495$~Gauss~\cite{Hou2019Experimental,Chang2025Hybrid}.
Subsystem A is encoded in two of the ground-state sublevels of the electron spin, specifically, $\ket{m_S=0}\equiv\ket{0}_\text{e}$ and $\ket{m_S=-1}\equiv\ket{1}_\text{e}$.
State initialization and readout are achieved through resonant optical transitions~\cite{Robledo2011Highfidelity}, achieving a single-shot readout fidelity of $(87.53\pm0.74)\%$ for both $\ket{0}_\text{e}$ and $\ket{1}_\text{e}$ (Supplementary Information). 
Coherent control of the electron spin is realized by using resonant microwave fields with a Rabi frequency of approximately $16.67$~MHz.
Subsystem B contains nine ${}^{13}\text{C}$ nuclear spins surrounding the NV center. Each nuclear spin qubit is encoded as $\ket{m_I=+1/2}\equiv\ket{0}_\text{n}$ and $\ket{m_I=-1/2}\equiv\ket{1}_\text{n}$. 
These nuclear spins exhibit long coherence times and weakly couple to the electron spin~\cite{Taminiau2012Detection,Taminiau2014Universal}.
They can be universally controlled using tailored microwave pulses applied to the electron spin~\cite{Taminiau2014Universal}.
As nuclear spins lack direct optical transitions, their initialization and readout are implemented by swapping their states with the electron spin using the universal gate set~\cite{Taminiau2014Universal}.

We obtain experimental trajectories following the scheme illustrated in Fig.~\ref{fig:2}\textbf{a}.
For the forward process, each experimental trial begins by initializing the ten spin qubits in $\rho^0$, a product of local Gibbs thermal states, via probabilistic sampling (Methods). The electron spin and the nuclear spin bath are at different temperatures. 
The system undergoes unitary evolution $U$, which is realized via the circuit shown in Fig.~\ref{fig:2}\textbf{b}, \textbf{c}. This unitary is followed by projective measurements of all spins in the Pauli-$Z$ basis. 
Ideal projective measurements yield outcomes and the system collapses onto a corresponding state $\rho^1$. The scheme requires $\rho^1$ to serve as the initial state for the next step. 
However, due to the destructive nature of the measurements in our system, the post-measurement state disagrees with the outcomes. 
To overcome this issue, we re-prepare the quantum state after each step based on the outcomes (Methods).
We implement $n=4$ steps and record the sequential measurement outcomes as a single forward trajectory, represented as a binary matrix of size $\left[10 \times 5\right]$. Backward trajectories are acquired in a similar fashion using the time-reversed protocol as mentioned above.
In total, we collect $45,000$ trajectories each for the forward and backward processes.

\begin{figure*}[t]
\includegraphics[width=\textwidth]{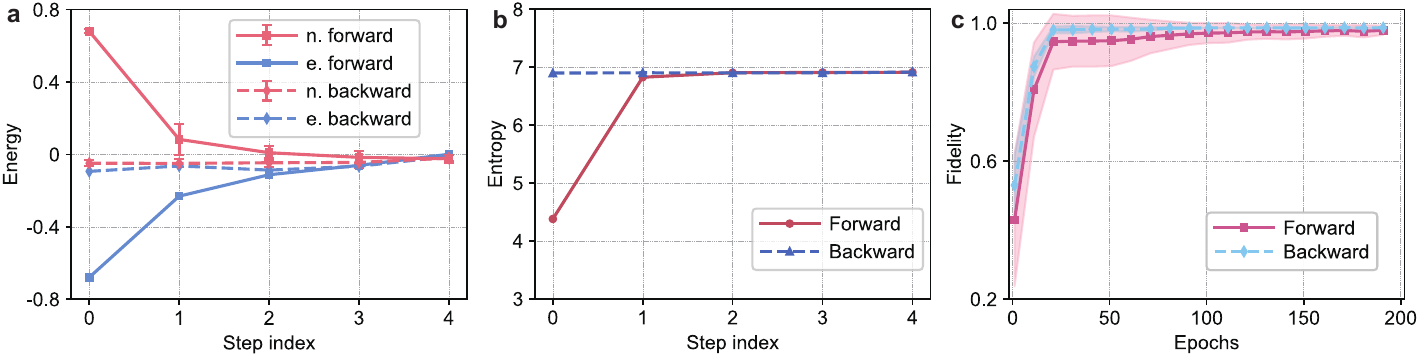}
\caption{
\textbf{Thermodynamic trajectories generated by a diffusion-based generative model.}
\textbf{a}, Energy of the electron spin (blue symbols)
and mean energy of the nine nuclear spins (red symbols), averaged over $90,000$ samples synthesized by the diffusion-based generative model, are plotted against the step index for both forward (squares) and backward (diamonds) processes. Error bars represent one standard deviation of the mean over the nine nuclear spins.
\textbf{b}, Plots of system entropy calculated from the trajectories generated by the diffusion-based model as a function of step index. 
Simulated data in \textbf{b},\textbf{c} is consistent with the experimental observations in Figs.~\ref{fig:2}\textbf{d}, \ref{fig:2}\textbf{e}. 
\textbf{c}, Fidelities between the generated and experimental states, averaged over all steps for the forward (red squares) and backward (blue diamonds) processes, are plotted as a function of training epochs (Methods). Error bars represent the standard deviation across time steps. 
}
\label{fig:4}
\end{figure*}

\vspace{.6cm}
\noindent\textbf{\large{}Machine learning the arrow of time} \\
We employ machine learning models to autonomously identify the arrow of time from experimental data. While the spin system under unitary evolution is intrinsically reversible, the arrow of time emerges through the stochasticity introduced by projective measurements (Fig.~\ref{fig:1}\textbf{a}).
This is revealed through statistical analysis of the experimental data as shown in Fig.~\ref{fig:2}\textbf{d}, \textbf{e}. During the forward process, the energy flows from the hotter nuclear spin qubits (Fig.~\ref{fig:2}\textbf{d}, red squares) to the colder electron spin qubit (Fig.~\ref{fig:2}\textbf{d}, blue squares), while the von Neumann entropy (Fig.~\ref{fig:2}\textbf{e}, red circles) monotonically increases and saturates at the end of the process. During the backward process, energies in both subsystems (Fig.~\ref{fig:2}\textbf{d}, red and blue diamonds) stay constant as the initial entropy (Fig.~\ref{fig:2}\textbf{e}, blue triangles) is already near the maximum.

We identify the arrow of time with two complementary approaches: an unsupervised clustering algorithm~\cite{cui2020introduction} and a supervised convolutional neural network (Methods).
In the unsupervised regime,  we apply the $k$-means clustering algorithm to the experimental trajectory dataset without access to temporal labels. The algorithm autonomously partitions the dataset into two distinct clusters (red and blue symbols in Fig.~\ref{fig:3}\textbf{a}) that correspond to the forward and backward processes, achieving a separation accuracy of approximately $90.61\%$. 
Furthermore, a convolutional neural network is trained using a labeled experimental dataset, which is randomly partitioned into a training set of $72,000$ samples and a test set of $18,000$ samples.
After training, the neural network achieves a classification accuracy of $92\%$ on the test set in distinguishing forward from backward trajectories. The binary cross-entropy losses on the training and test sets converge to a steady value (Fig.~\ref{fig:3}\textbf{b}).
These results demonstrate that the temporal asymmetry lying in experimental trajectories can be effectively extracted by machine learning models.
The fact that the accuracy does not reach $100\%$ is mainly attributed to the stochastic nature of microscopic thermodynamics in a finite size system and experimental imperfections.

To validate the learning results (Fig.~\ref{fig:3}\textbf{b}) originating from the arrow of time, we re-train the convolutional neural network using synthetic trajectories. These trajectories are generated from numerical simulations which only contain unitary evolution and no intermediate projective measurements applied (Methods). In this coherent regime, the classification accuracy of the same neural network architecture drops to the random-guessing baseline of around $50\%$ (Fig.~\ref{fig:3}\textbf{c}).
This sharp contrast provides compelling evidence that the arrow of time identified by the machine learning model is induced by projective measurements, rather than any inherent details of unitary dynamics.
We remark that both models operate in a purely data-driven manner, unassisted by predefined physical formulas or explicit knowledge of the system evolution.
In addition, experimental imperfections, such as gate errors and readout noise, inevitably cause the actual dynamics to deviate from the ideal protocol engineered in Fig.~\ref{fig:2}\textbf{a}. However, the machine learning models successfully identify the arrow of time despite these deviations.

\vspace{.6cm}
\noindent\textbf{\large{}Generative modeling of the arrow of time} \\
We next investigate the capability of generative models to capture the core signature of the arrow of time.
Unlike classification models, which simply map inputs to labels, generative models must capture essential features from noisy experimental data to produce data consistent with outputs from physical systems. 
We employ a diffusion-based generative model~\cite{Sohl-Dickstein2015Deep,Ho2020Denoising}, whose network architecture is shown in Fig.~\ref{fig:1}\textbf{d}.
Crucially, the generative model is constructed without any explicit knowledge of the experimental system and physical principles.
The complete generation pipeline is outlined in Fig.~\ref{fig:1}\textbf{e}. The model iteratively denoises Gaussian white noise to construct physically plausible evolution trajectories. We generate $90,000$ synthetic trajectories, and extract the energies (Fig.~\ref{fig:4}\textbf{a}) and the von Neumann entropy (Fig.~\ref{fig:4}\textbf{b}) as a function of step.  These statistical results of the generated trajectories well match the experimental observations shown in Fig.~\ref{fig:2}\textbf{d} and \ref{fig:2}\textbf{e}. We also estimate the fidelity between the projected physical state and the generated state to be approximately $0.982$ after training (Fig.~\ref{fig:4}\textbf{c}, Methods).
These results validate that the machine learning model effectively captures the core signature of the arrow of time.

\vspace{.6cm}
\noindent\textbf{\large{}Conclusion and outlook} \\
In summary, our work employs machine learning techniques to identify the arrow of time in individual trajectories from a quantum system.
By training on experimental data, we show that discriminative models, clustering, and convolutional neural networks can effectively learn the temporal irreversibility induced by projective measurements with high accuracy, limited by the intrinsic overlap between the two distributions. 
The diffusion-based generative model without prior physical knowledge autonomously captures the underlying thermodynamic principles directly from noisy experimental data. 

The machine learning methodology presented here is inherently scalable. While this work employs the evolution of ten qubits as a proof-of-principle demonstration, the framework can be readily applied to characterize the thermodynamic evolution of larger-scale quantum systems.
Looking forward, a natural extension is to apply this framework to open quantum systems, where measurement dynamics and environmental dissipation coexist, leading to complex non-equilibrium behavior. Furthermore, investigating the link between the arrow of time and quantum entanglement dynamics represents a future direction. These studies may deepen our understanding of microscopic thermodynamics.

\vspace{.5cm}
\noindent\textbf{Acknowledgement} \\
We thank M. Hafezi and Wenjie Jiang for helpful discussions.
We acknowledge support from the Quantum Science and Technology-National Science and Technology Major Project (Grant No.~2021ZD0302203), the National Natural Science Foundation of China (grant No.~T2225008), the Shanghai Qi Zhi Institute Innovation Program SQZ202318, and the National Key R\&D Program of China (grant no.~2023YFB4502600).
W.L., D.-L.D., and P.-Y.H. are supported in addition by Tsinghua University Dushi Program.
P.-X.S. acknowledges support from the European Union's Horizon Europe research and innovation programme under the Marie Skłodowska-Curie Grant Agreement No.~101180589 (SymPhysAI), the National Science Centre (Poland) OPUS Grant No.~2021/41/B/ST3/04475, and the Foundation for Polish Science project MagTop (No.~FENG.02.01-IP.05-0028/23) co-financed by the European Union from the funds of Priority 2 of the European Funds for a Smart Economy Program 2021–2027 (FENG).
Views and opinions expressed are however those of the author(s) only and do not necessarily reflect those of the European Union or the European Research Executive Agency. Neither the European Union nor the granting authority can be held responsible for them.

\makeatother
\bibliography{EMLAT}

\renewcommand{\figurename}{Extended Fig.}
\setcounter{figure}{0}

\clearpage
\noindent\textbf{\large{}Appendix}

\vspace{.2cm}
\noindent\textbf{Diamond sample and control system} \\
Experiments are performed on a high-purity synthetic diamond in a cryogenic environment (approximately $7$~K). A $532$-nm laser prepares the NV center in the negative charge state.  Two red lasers at $637$-nm, labeled $E_{1,2}$ and $E_y$, drive transitions $\ket{\pm 1}_\text{e}\rightarrow\ket{E_{1,2}}_\text{e}$ and $\ket{0}_\text{e}\rightarrow\ket{E_y}_\text{e}$, respectively (Fig.~\ref{fig:1}\textbf{b}). The $E_{1,2}$ laser initializes the electron spin in $|0\rangle_\text{e}$. The $E_y$ laser is used to read out the electron spin state: $|0\rangle_\text{e}$ acts as a bright state while $|1\rangle_\text{e}$ as a dark state.  
More experimental details are provided in the Supplementary Information.

\vspace{.2cm}
\noindent\textbf{Experimental data collection procedure} \\ 
Each experimental trajectory begins with all spins prepared in Gibbs thermal states, which is obtained through probabilistic sampling from a prescribed distribution of pure states. 
We adopt a state-repreparation scheme, in which the quantum state is re-initialized after each measurement based on its outcome. 
Concretely, we perform an ensemble of experiments, each of which begins with a distinct pure state, evolves under unitary $U$, and ends with projective measurements. We record initial pure states, measurement outcomes, and their corresponding projected states, thereby forming a large dataset. Full evolution trajectories are then obtained by assembling the single-step data randomly sampled from the dataset. 
This entire procedure is equivalent to the aforementioned procedure based on probabilistic sampling and state re-preparation. More details are provided in the Supplementary Information.

\vspace{.2cm}
\noindent\textbf{Numerically generated trajectories} \\ 
To verify that the machine learning results in Fig.~\ref{fig:3}\textbf{b} arise from the arrow of time introduced by projective measurements, we generate trajectories via numerical simulations of unitary evolution with no measurements.

Numerical simulations start with an initial Gibbs thermal state prepared at the same temperatures as those used in the experiment. The system then evolves under the unitary operator $\prod_{i=N}^1\mathcal{U}_i$ for multiple steps. After each step, binary outcomes ($0$ or $1$) are probabilistically sampled for each qubit based on its Pauli-\textit{Z} expectation value, collectively forming the forward evolution trajectory. Reverse evolution trajectories are generated similarly under the time-reversed unitary $\prod_{i=1}^N\mathcal{U}_i^{\dagger}$. Crucially, this numerical simulation assumes that measurements only extract information from the system without collapsing the system---a scenario that cannot be implemented experimentally.

\begin{figure}[htbp]
\centering
\includegraphics[width=\linewidth]{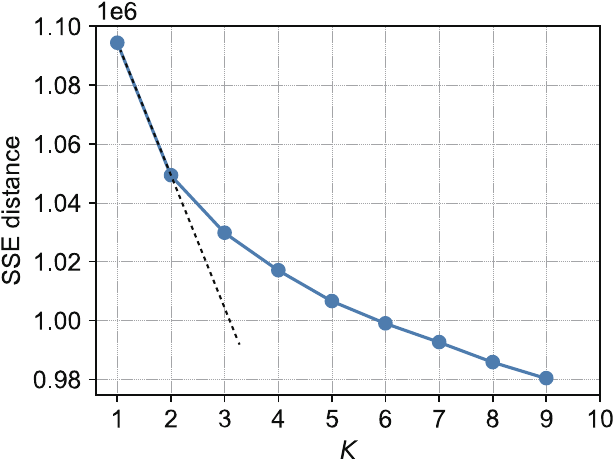}
\caption{
\textbf{The elbow method for determining the optimal number of clusters.}
The elbow method is employed to determine the optimal number of clusters $K$, using the within-cluster sum of squared Euclidean (SSE) distances (the sum of squared Euclidean distances from each sample to its cluster centroid). 
As $K$ increases, the SSE initially decreases rapidly and then gradually slows down, forming an elbow-like shape. The $K$ at this elbow balances tight intra-cluster grouping and avoids over-clustering. Here, the optimal number of clusters is determined to be $K=2$. The black dotted line serves as a guide to the eye, highlighting the sharp transition in the slope at this optimal value.
}
\label{fig:ext1}
\end{figure}

\vspace{.2cm}
\noindent\textbf{Clustering algorithm} \\
The $k$-means algorithm is a widely used unsupervised clustering method that partitions data into a predefined number of clusters by minimizing the within-cluster sum of squared Euclidean (SSE) distances. The algorithm iterates two core steps: assigning each sample to the nearest centroid, and updating each centroid to be the mean of the samples in its cluster.
To address potential empty clusters during iteration, the algorithm randomly reinitializes empty centroids using the input data. The iterative process terminates when centroid updates fall below a specified tolerance threshold or reach the maximum number of iterations, ensuring both efficiency and convergence (Algorithm~\ref{cluster}). We employ the elbow method to determine the optimal number of clusters (Extended Fig.~\ref{fig:ext1}).

\begin{algorithm}[htbp]
\label{cluster}
\DontPrintSemicolon
\KwIn{Dataset $X = \{x_1, \dots, x_{S}\}$, number of clusters $K$, maximum iterations $M$, tolerance $\tau$}
\KwOut{Centroids $C = \{c_1, \dots, c_K\}$, cluster labels $L = \{l_1, \dots, l_S\}$}

\BlankLine

Initialize $C$ by randomly selecting $K$ distinct points from $X$\;

\BlankLine

\For{$t \gets 1$ \KwTo $M$}{
    \tcp{Assignment step}
    \For{$i \gets 1$ \KwTo $N$}{
        $l_i \gets \arg\min_{k \in \{1,\dots,K\}} \|x_i - c_k\|_2^2$\;
    }
    
    \tcp{Update step}
    \For{$k \gets 1$ \KwTo $K$}{
        $X_k \gets \{x_i \in X \mid l_i = k\}$\;
        
        \eIf{$X_k \neq \emptyset$}{
            $c_k^{\text{new}} \gets \frac{1}{|X_k|} \sum_{x \in X_k} x$\;
        }{
            $c_k^{\text{new}} \gets $ randomly selected point from $X$\;
        }
    }

    \tcp{Convergence check}
    \If{$\sum_{k=1}^K \|c_k^{\textup{new}} - c_k\|_2^2 < \tau$}{
        \textbf{break}\;
    }
    
    $C \gets C_{\text{new}}$\;
}

\Return $C, L$\;

\caption{Clustering algorithm: $k$-means}
\end{algorithm}

\vspace{.2cm}
\noindent\textbf{Convolutional neural network} \\
The convolutional neural network (CNN) architecture shown in Fig.~\ref{fig:1}c processes input tensors of shape $(B, T, N)$, where $B$ is the batch size, $T$ is the number of time steps, and $N$ is the number of qubits. To enable 2D convolution, a channel dimension is introduced, reshaping the input to $(B, 1, T, N)$. A convolutional layer with a $(2, 1)$ kernel, no padding, and two output channels is applied, followed by a ReLU activation function. This operation effectively captures temporal correlations between adjacent time steps, yielding a tensor of shape $(B, 2, T-1, N)$. The output is then flattened into a vector of size $2(T-1)N$ and fed into a fully connected layer with $\lfloor TN/2 \rfloor$ neurons, followed by another ReLU activation. To improve generalization, batch normalization and a dropout layer with a rate of $0.2$ are applied. Finally, a fully connected layer projects the extracted features to a 1D space, and a sigmoid activation function outputs a scalar probability of shape $(B, 1)$ to determine the temporal direction of the trajectory.

\vspace{.2cm}
\noindent\textbf{Generative model based on diffusion} \\
The diffusion model (Fig.~\ref{fig:1}\textbf{d}, \textbf{e}) is implemented through a systematic process of data encoding, forward noise addition, and reverse denoising. 
First, input trajectories of shape $(S, T, N)$, where $S$ is the number of samples, are flattened to a dimension of $T \times N$.
The binary states $\{0,1\}$ are mapped to $\{-1,1\}$, and the temporal direction labels are encoded as one-hot vectors. 

For the forward diffusion process, we adopt a linear noise schedule $\beta_l = \beta_{\text{start}} + l(\beta_{\text{end}} - \beta_{\text{start}})/L$, utilizing $L=1000$ diffusion timesteps with $\beta_{\text{start}}=10^{-4}$ and $\beta_{\text{end}}=0.02$. The noisy state at diffusion step $l$ is generated via $x_l = \sqrt{\bar{\alpha}_l} x_0 + \sqrt{1-\bar{\alpha}_l} \epsilon$, where $\epsilon \sim \mathcal{N}(0, I)$ is the added Gaussian noise and $\bar{\alpha}_l$ represents the cumulative product of $1-\beta_l$. 
The core denoising architecture is a multilayer perceptron with a hidden dimension of 128. It generates sinusoidal embeddings for the diffusion timesteps and 2D embeddings for the direction labels using sequential linear and SiLU activation layers. These are concatenated with the projected input features to form a $3 \times 128$-dimensional representation. A three-layer linear network with SiLU activations and Xavier initialization then predicts the added noise $\epsilon$. The model is trained by minimizing the mean squared error between the predicted and true noise. 

During the generation phase, synthetic trajectories are produced from pure Gaussian noise via iterative reverse denoising steps. The continuous-valued outputs are binarized back to $\{-1,1\}$, then mapped back to $\{0,1\}$, and finally reshaped to the original $(S, T, N)$ tensor format.

To evaluate the generative performance, we quantify the fidelity between the experimental density matrix $\rho_{\text{exp}}$ and the generated density matrix $\rho_{\text{gen}}$. The fidelity, averaged over all $n+1$ trajectory time steps, is calculated as
$$ \text{Fidelity} = \frac{1}{n+1} \sum_{k=0}^{n} \left( \operatorname{Tr}\left[ \sqrt{\sqrt{\rho_{\text{exp}}^{(k)}} \rho_{\text{gen}}^{(k)}\sqrt{\rho_{\text{exp}}^{(k)}}} \right] \right)^2, $$
where the superscript $(k)$ denotes the state at the $k$-th time step. The estimated fidelities are presented in Fig.~\ref{fig:4}c.

\clearpage
\newpage 
\onecolumngrid

\newcommand{\mcol}{\multicolumn{2}{c}}
\newcommand{\cfill}{\cellcolor[HTML]{44cef6}}
\newcommand{\cfillo}{\cellcolor[HTML]{f05654}}
\setcounter{MaxMatrixCols}{10}
\hypersetup{urlcolor=blue}

\xdef\presupfigures{\arabic{figure}}
\newcommand{\figpath}{./figures}

\setcounter{figure}{0}
\setcounter{table}{0}
\renewcommand{\theequation}{S\arabic{equation}}
\renewcommand{\thefigure}{S\arabic{figure}}
\renewcommand{\thetable}{S\arabic{table}}
\renewcommand{\figurename}{Fig.}
\renewcommand{\tablename}{Tab.}

\begin{center} 
{\large \bf Supplementary Information for \\ ``Machine learning the arrow of time in solid-state spins''}
\end{center} 

\maketitle

\section{Theoretical framework}
We consider a quantum thermodynamic model involving two subsystems: subsystem A (a single qubit) and subsystem B (a bath of $N$ qubits). The Hamiltonian for subsystem A is $H_{\text{A}} = Z_{\text{A}}$, and $H_{\text{B}} = \sum_{i=1}^N H_{\text{B}}^i = \sum_{i=1}^N Z_{i}$ for subsystem B, where $Z$ denotes the Pauli-$Z$ operator.
Initially, both subsystems are prepared in independent thermal Gibbs states. The joint state is described by: 
\begin{equation}
    \rho_{\text{AB}}^0 = \rho_{\text{A}}^0 \otimes \rho_{\text{B}}^0,
\end{equation}
where $\rho_{\text{A}}^0 = e^{-\beta_{\text{A}} H_{\text{A}}}/\mathcal{Z}_{\text{A}}$ and $\rho_{\text{B}}^0 = \bigotimes_{i=1}^N \left[e^{-\beta_i H_{\text{B}}^i}/\mathcal{Z}_i\right]$.
The superscript on $\rho$ indicates different time steps.
Here, $\beta_j$ represents the inverse temperature and $\mathcal{Z}_j = \operatorname{Tr}\left[\exp\left(-\beta_j H_j\right)\right]$ is the partition function, where $j = \text{A}\ \text{or}\ j\in \left\{1,...,N\right\}$.
The interaction between the subsystems is engineered via a unitary evolution operator defined as:
\begin{equation}
    U = \prod_{i=N}^1 \mathcal{U}_{i} = \prod_{i=N}^1 \exp\left[-\mathrm{i}\left(X_\text{A} X_{\text{i}} + Y_\text{A} Y_{i}\right)\Delta t\right],
    \label{eq:interaction_operator}
\end{equation}
where $\mathcal{U}_{i}$ generates the exchange interaction between subsystem A and the $i$-th qubit of subsystem B. In our demonstration, this model is realized using a nitrogen-vacancy (NV) center, where the electron spin acts as subsystem A (${X}_\text{A},{Y}_\text{A},{Z}_\text{A}={X}_\text{e},{Y}_\text{e},{Z}_\text{e}$) and ten of the surrounding ${}^{13}\text{C}$ nuclear spins act as subsystem B (see Fig.~1\textbf{b} in Main text).

\subsection{Arrow of time and entropy production}

An isolated quantum system evolves under the system Hamiltonian, $\rho(t) = U(t) \rho(0) U^\dagger(t)$, where $U(t)$ is unitary generated by the Hamiltonian. Unitary evolution conserves von Neumann entropy and is reversible via its Hermitian conjugate $U^\dagger(t)$, giving rise to time-reversal symmetry. However, the introduction of projective measurements breaks this time-reversal symmetry, defining an arrow of time through entropy production~\cite{Nielsen2010Quantum}. In the protocol, we consider forward and reverse processes consisting of multiple unitary evolution and projective measurements:
\begin{itemize}
\item Forward Process: The forward thermodynamic process consists of $n$ steps. In the $k$-th step, subsystem A interacts with subsystem B via $U$, followed by a projective measurement of the total system in the complete $Z$-basis $\{P_a \mid a=1,\dots,2^{N+1}\}$, where projectors $P_1=\ket{00\dots 00}\bra{00\dots 00}$, $P_2=\ket{00\dots 01}\bra{00\dots 01}$, $\dots$, $P_{2^{N+1}}=\ket{11\dots 11}\bra{11\dots 11}$.
The initial state of the $k$-th step $\rho^{k-1}_{\text{AB}}$ evolves to $\rho^{(k-1)\mapsto k}_{\text{AB}} = U \rho^{k-1}_{\text{AB}}U^{\dagger}$. Then, the measurement collapses the system, yielding the state
\begin{equation}
\rho^{k}_{\text{AB}} = \sum_a P_a (\rho^{(k-1)\mapsto k}_{\text{AB}}) P_a = \sum_a P_a U \rho^{k-1}_{\text{AB}} U^{\dagger} P_a.
\end{equation}
This post-measurement state $\rho^{k}_{\text{AB}}$ is diagonal in the Pauli-$Z$ basis. Since projective measurements generally increase entropy, we have $S(\rho^{k}_{\text{AB}}) \geq S(\rho^{k-1\mapsto k}_{\text{AB}}) = S(\rho^{k-1}_{\text{AB}})$, where $S(\rho^{k}_{\text{AB}}) = S(\rho^{k-1\mapsto k}_{\text{AB}})$ holds if and only if the state $\rho^{k-1\mapsto k}_{\text{AB}}$ is already diagonal in the computational basis. Here, von Neumann entropy of state $\rho$ is evaluated as $S(\rho) = -\operatorname{Tr}\left[\rho \ln (\rho)\right]$. Repeating this for $n$ steps yields the forward state trajectory $\boldsymbol{\rho}^{F} = \{ \rho^{k}_{\mathrm{AB}}\}_{k=0}^n$.
\item Reverse Process: The reverse process begins with the final state of the forward trajectory, $\sigma^0_{\text{AB}} = \rho^{n}_{\text{AB}}$. In each step, the system interacts via the inverse operator $U^{\dagger}$ and is subsequently measured in the Pauli-$Z$ basis. The state evolution is given by:
\begin{equation}
    \sigma^{k}_{\text{AB}} = \sum_a P_a U^{\dagger} (\sigma^{k-1}_{\text{AB}}) U P_a.
\end{equation}
This generates the reverse state trajectory $\boldsymbol{\sigma}^{R} = \{ \sigma^{k}_{\mathrm{AB}}\}_{k=0}^n$. We define the ``backward trajectory'' as the time-reversal of this sequence: $\overline{\boldsymbol{\sigma}^{R}} = \{ \sigma^{n-k}_{\mathrm{AB}}\}_{k=0}^n$.
In this protocol, $\boldsymbol{\rho}^{F} \neq \overline{\boldsymbol{\sigma}^{R}}$. The entropy strictly increases along $\boldsymbol{\rho}^{F}$ but would appear to decrease for the time-reversed sequence $\overline{\boldsymbol{\sigma}^{R}}$, indicating the irreversibility of the thermodynamic process.
\end{itemize}

\subsection{Thermodynamic interpretation: heat and energy}

In macroscopic thermodynamics, heat is defined as the energy exchanged between two large systems at different temperatures. This concept has been extended to small quantum systems initially prepared in thermal Gibbs states~\cite{Micadei2019Reversing}.
In our protocol, the interaction operator $\mathcal{U}_{i}$ commutes with the total Hamiltonian of the two interacting spins, $[H_{\text{A}} + H_{\text{B}}^i, \mathcal{U}_{i}] = 0$. Consequently, the unitary operation $\mathcal{U}_i$ performs no work on the whole system. Furthermore, since the projective measurement operators $P_a$ commute with $H_{\text{A}}$ and $H_{\text{B}}$, the measurements also conserve the energy of the subsystems (see Proposition 1 in \ref{subsection:proof}).
Thus, the total energy remains constant during the $k$-th step. Any local energy change is attributed to heat exchange between subsystems. The heat absorbed by subsystem $j$ ($j \in \{\text{A, B}\}$) is defined as:
\begin{equation}
    Q^{k}_{j} = \Delta E^{k}_{j} = E_{j}^k - E_{j}^{k-1},
\end{equation}
satisfying the conservation law $Q^{k}_{\text{A}} = -Q^{k}_{\text{B}}$.

To understand the direction of this heat flow, we examine the microscopic interaction in detail. In the $k$-th step, the total system is initially in a diagonal state $\rho^{k-1}_{\text{AB}}$. Subsystem A interacts sequentially with each qubit in subsystem B via the unitary $\mathcal{U}_{i}$ ($i=1, \dots, N$).
After the application of unitaries involving the first $i-1$ qubits, the reduced joint state of the qubit in subsystem A and the $i$-th qubit in subsystem B is described as $\rho_{\text{A},i} = \operatorname{Tr}_{(\overline{\text{A},i})} [\mathcal{U}_{i-1} \dots \mathcal{U}_{1} \rho^{k-1}_{\text{AB}} \mathcal{U}_{1}^{\dagger} \dots \mathcal{U}_{i-1}^{\dagger}]$, where $(\overline{\text{A},i})$ denotes the trace over all qubits except subsystem A and qubit $i$. 
Crucially,  this state remains diagonal in the measurement basis (see Proposition 2 in \ref{subsection:proof}). The local temperatures of both subsystem A and qubit $i$ remain well-defined. 

Under these conditions, the interaction $\mathcal{U}_{i}$ dictates that heat flows naturally from the hotter qubit to the colder qubit. Specifically, if qubit $i$ is colder than subsystem A (i.e., $\beta_{\text{A}} \leq \beta_{i}$), it absorbs heat, resulting in $Q^{k}_{i} > 0$ (see Proposition 3 in \ref{subsection:proof}).
This mechanism establishes a link between the arrow of time and thermodynamic heat transfer. In the forward process, assuming subsystem A is initially colder than subsystem B, it continuously absorbs energy across the sequential interactions, leading to an increasing energy profile in the forward trajectory $\boldsymbol{\rho}^{F}$. Conversely, the time-reversed backward trajectory $\overline{\boldsymbol{\sigma}^{R}}$ would depict subsystem A releasing energy to a hotter bath, a process that violates the second law of thermodynamics.

\subsection{Propositions}
\label{subsection:proof}
\textbf{Proposition 1:} As the projective measurement operators $P_a$ commute with $H_{\text{A}}$ and $H_{\text{B}}$, the measurements conserve the energy of the subsystems, i.e., $E_{\text{A}}^{(k-1)\mapsto k} = E_{\text{A}}^{k} $ and $E_{\text{B}}^{(k-1)\mapsto k} = E_{\text{B}}^{k} $.

\textit{Proof.}
We prove $E_{\text{A}}^{k-1\mapsto k} = E_{\text{A}}^{k} $ as follows:
\begin{equation}
\begin{aligned}
    E_{\text{A}}^k &= \operatorname{Tr}\left[ (H_\text{A} \otimes \mathbb{I}_\text{B}) \rho^k_{\text{AB}} \right] \\
    &= \sum_a \operatorname{Tr}\left[ (H_\text{A} \otimes \mathbb{I}_\text{B}) P_a (U \rho^{k-1}_{\text{AB}} U^{\dagger}) P_a \right] \\
    &= \operatorname{Tr}\left[ (H_\text{A} \otimes \mathbb{I}_\text{B}) (U \rho^{k-1}_{\text{AB}} U^{\dagger}) \sum_a P_a^2 \right] \text{(since}~[P_a, U]=0~\text{and}~[P_a, \rho_{AB}^{k-1}]=0 \text{)} \\
    &= \operatorname{Tr}\left[ H_\text{A} \rho^{k-1\mapsto k}_{\text{A}} \right] \\
    &= E_{\text{A}}^{k-1\mapsto k},
\end{aligned}
\end{equation}

where $\rho^k_{\text{A}} = \operatorname{Tr}_{\text{B}}(\rho^k_{\text{AB}})$, $\rho^{k-1\mapsto k}_{\text{A}} = \operatorname{Tr}_{\text{B}}(\rho^{k-1\mapsto k}_{\text{AB}})$.
In a similar way, $E_{\text{B}}^{k-1\mapsto k} = E_{\text{B}}^{k}$ can be proved. \hfill $\blacksquare$

\textbf{Proposition 2:} The partial trace $\operatorname{Tr}_{(\overline{\text{A},i})} \left[\mathcal{U}_{i-1} \dots \mathcal{U}_{1} \rho^{k-1}_{\text{AB}} \mathcal{U}_{1}^\dagger \dots \mathcal{U}_{i-1}^{\dagger}\right]$ is diagonal, where $\rho^{k-1}_{\text{AB}}$ is diagonal and $(\overline{\text{A},i})$ denotes the trace over all qubits except subsystem A and qubit $i$.

\textit{Proof.}
Consider a $Z$-basis state $|Z_a\rangle = |l_{\text{A}}, l_1, \dots, l_N \rangle$, with $l_k \in \{0,1\}$. We express the evolved state after applying the sequence of unitaries  $\mathcal{U}_{1}, \dots, \mathcal{U}_{i-1}$ as
\begin{equation}
    \mathcal{U}_{i-1} \dots \mathcal{U}_{1} |Z_a\rangle = \sum_{\mathbf{l}'} C_{\mathbf{l}'} |l'_{\text{A}}, l'_1, \dots, l'_{i-1}\rangle \otimes |l_{i}, \dots, l_N \rangle,
    \label{eq:expansion}
\end{equation}
where the coefficients $C_{\mathbf{l}'}$ are non-zero only when $l'_{\text{A}} + \sum_{m=1}^{i-1} l'_m = l_{\text{A}} + \sum_{m=1}^{i-1} l_m$. This is because each unitary preserves the total excitation number (energy) of the involved qubits. 

Now consider two distinct basis states $\{\mathbf{l}'\}$ and $\{\mathbf{l}''\}$. For a non-zero off-diagonal element to survive the partial trace over indices $1 \dots i-1$, it requires $\{l'_1, \dots, l'_{i-1}\} = \{l''_1, \dots, l''_{i-1}\}$.
Combined with the energy conservation constraint, the quantum numbers of system A in these two bases must also be identical, i.e. $l'_{\text{A}} = l''_{\text{A}}$. Therefore, no off-diagonal terms survive. Since $\rho^{k-1}_{\text{AB}}$ is a mixture of $|Z_a\rangle\langle Z_a|$, the resulting reduced state is diagonal. \hfill $\blacksquare$

\textbf{Proposition 3:} For an arbitrary diagonal two-qubit state $\rho_{\text{A},i}$, the qubits are locally in thermal states, and the interaction $\mathcal{U}_{i}$ induces heat flow from hot to cold.

\textit{Proof.}
Any diagonal two-qubit state can be decomposed as:
\begin{equation}
    \rho_{\text{A},i} = \rho_{\text{A}} \otimes \rho_{i} + \alpha K,
\end{equation}
where $K = |00\rangle\langle 00| + |11\rangle\langle 11| - |01\rangle\langle 01| - |10\rangle\langle 10|$, and $\rho_{\text{A}}$ and $\rho_{i}$ are local thermal states. 
Note that $\operatorname{Tr}_{\text{A}}(K) = \operatorname{Tr}_{i}(K) = 0$; first, consider the uncorrelated case ($\alpha=0$). The heat exchange satisfies the relation based on mutual information change $\Delta I(\text{A}:i)$~\cite{Micadei2019Reversing}:
\begin{equation}
    \beta_{\text{A}} Q_{\text{A}} + \beta_{i} Q_{i} \geq \Delta I(\text{A}:i),
\end{equation}
where ($\beta_A$, $Q_A$) and ($\beta_i$, $Q_i$) are the inverse of temperature and absorbed heat of qubits $A$ and $i$, respectively. 
Since the initial mutual information is zero and correlations generally increase ($\Delta I \geq 0$), and given $Q_{\text{A}} = -Q_{i}$, we obtain:
\begin{equation}
    Q_i (\beta_{i} - \beta_{\text{A}}) \geq 0.
\end{equation}
This inequality implies that if $\beta_i \geq \beta_{\text{A}}$ (qubit $i$ is colder), then $Q_i \geq 0$ (it absorbs heat). 
For the correlated case ($\alpha \neq 0$), since $[\mathcal{U}_{i}, K] = 0$, the correlation term $K$ is invariant under the interaction and does not contribute to energy transfer. Thus, the direction of heat flow is dictated solely by the local thermal parts, preserving the ``hot to cold'' principle. \hfill $\blacksquare$

\section{Experimental setup}

\subsection{Diamond sample}

A high-purity synthetic diamond sample with a natural abundance of ${}^{13}\mathrm{C}$ isotope of approximately $1.1\%$ is used in the experiment~\cite{Hou2019Experimental}.
Experiments are performed using an optical confocal system equipped with an objective lens with a numerical aperture of approximately $0.9$.
The diamond sample is mounted on a three-dimensional piezoelectric positioner inside a closed-cycle cryostat at a temperature of approximately $7$~K.
To enhance fluorescence collection efficiency, a solid-immersion lens (SIL) is fabricated directly on the diamond surface at the location of a single NV center.
A permanent magnet placed outside the cryostat applies a magnetic field of approximately $495$~G along the NV axis.
Additionally, a gold stripline is fabricated around the SIL to deliver microwave fields, enabling control of the electron spin of the NV center (Fig.~\ref{fig:system}).

\subsection{Control system}

\begin{figure}[t]
\centering
\includegraphics[width=\textwidth]{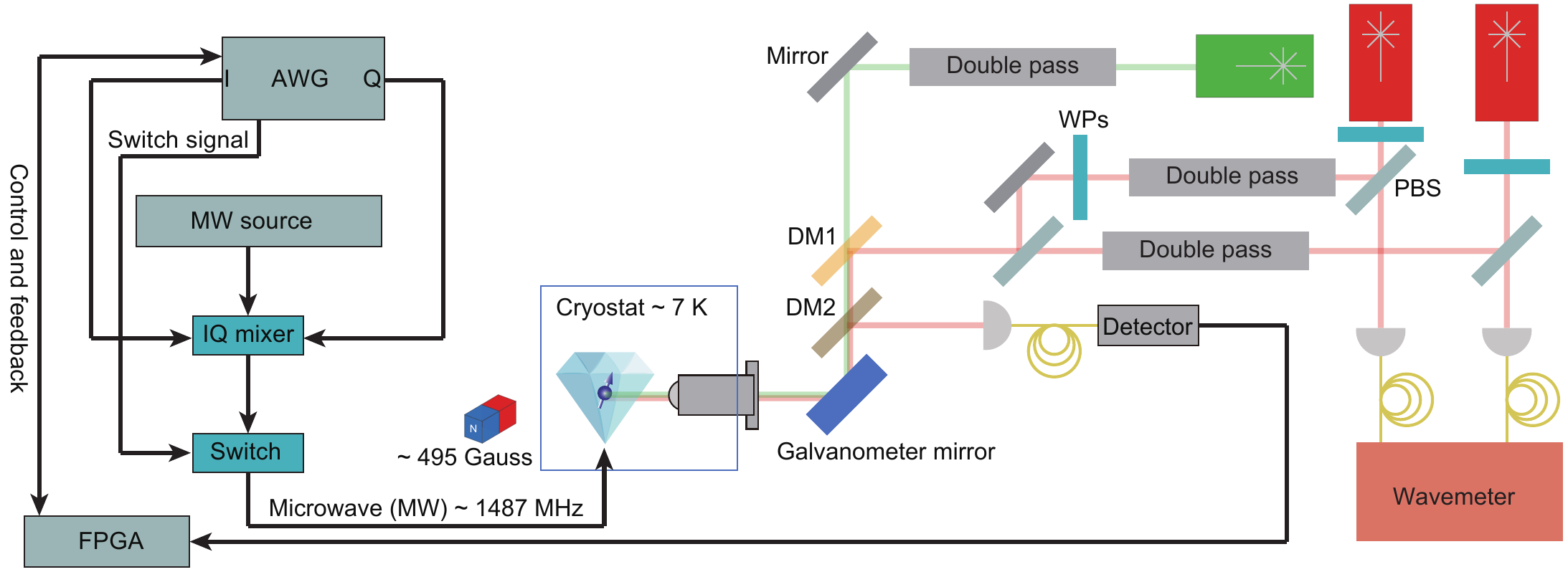}
\caption{
\textbf{Sketch of experimental setup.}
The schematic shows the control and optical systems. The control system generates programmable microwave pulses, which are produced by combining the outputs of an arbitrary wave generator (AWG) and a microwave (MW) source via an IQ mixer, and controlled by an RF switch. A field-programmable gate array (FPGA) device is used to count the NV fluorescence as experimental outcomes and is able to trigger the AWG based on the NV outcomes in real time.
The optical system consists of three lasers. One green laser at 532\,nm is free-running, and two red lasers at 637\,nm are frequency locked to a commercial wavemeter. These laser beams are switched on and off by acoustic-optic modulators in double-pass configurations, combined by dichroic mirrors (DMs), and directed onto the diamond sample by a galvanometer mirror. WP: wave plates, PBS: polarizing beam splitters.
}
\label{fig:system}
\end{figure}

All experimental sequences are controlled by a commercial arbitrary waveform generator (AWG) coupled with a field-programmable gate array (FPGA) device (Fig.~\ref{fig:system}). The microwave field is generated by mixing the outputs of a microwave signal generator and the AWG using an IQ mixer. Meanwhile, the AWG also controls the laser switch. Fluorescence photons emitted by the NV center are detected via a single-photon detector (SPD), which converts the optical signal into an electrical signal. 
Moreover, the same AWG regulates the length of the photon-counting window to optimize single-shot readout fidelity (Fig.~\ref{fig:readoutFidelity}). Furthermore, we implement a real-time feedback module for photon counting by integrating the FPGA development board with the AWG. Based on the counting results, the FPGA board sends an analog signal to the AWG to control experimental sequences in real time.~\cite{Chang2025Hybrid}.

\subsection{Optical system}

Three laser beams are used in the experiment: a 532\,nm laser initializes the NV center charge state, and two 637\,nm lasers serve for electron spin polarization into $\ket{0}_\text{e}$ and for electron spin readout (Fig.~\ref{fig:system}). The two red lasers, denoted as $E_{1,2}$ and $E_y$, drive the transitions $\ket{\pm 1}_\text{e}\rightarrow\ket{E_{1,2}}_\text{e}$ and $\ket{0}_\text{e}\rightarrow\ket{E_y}_\text{e}$. With the AWG, these lasers also detect the nitrogen nuclear spin state and enable ${}^{14}\text{N}$ nuclear spin polarization based on measurements.

The directions of all three beams are controlled by a scanning galvanometer mirror combined with a $4f$ optical system, enabling precise positioning at the NV center. Before entering the $4f$ system, each red laser beam is split by a polarizing beam splitter (PBS). One portion is coupled into a single-mode fiber and sent to a wavelength meter for frequency stabilization, while the other is directed onto the NV center.

\begin{figure}[t]
\centering
\includegraphics[width=0.7\linewidth]{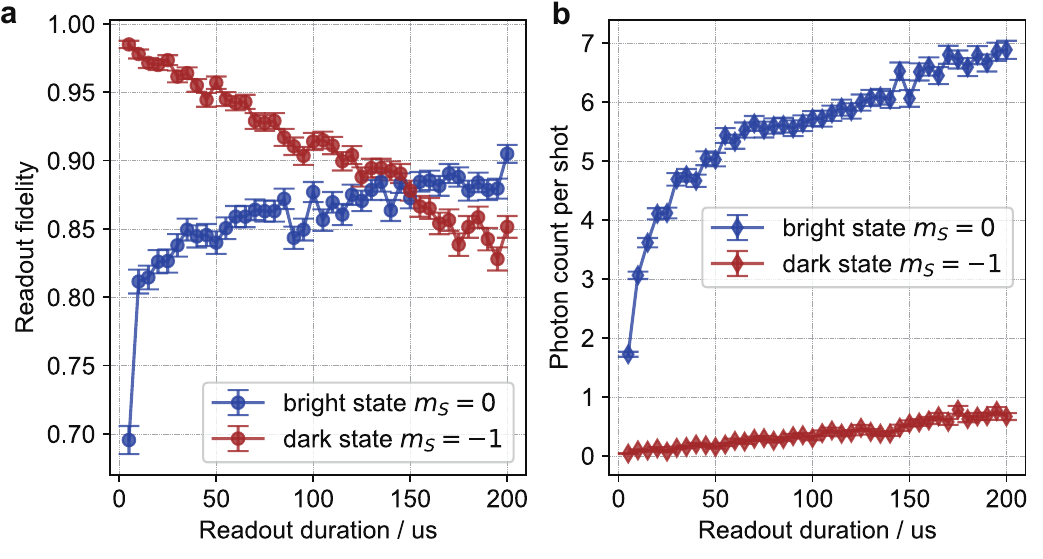}
\caption{
\textbf{Readout fidelity of the NV center electron spin over readout duration time.}
\textbf{a}, Single-shot readout fidelities of the bright state (blue circles) and the dark state (red circles) are plotted as a function of readout duration.
\textbf{b}, Plot of the average photon counts for the bright state (blue diamonds) and dark state (red diamonds) as a function of readout duration. Error bars represent one standard deviation.
}
\label{fig:readoutFidelity}
\end{figure}

Laser switching is implemented via an acousto-optic modulator (AOM) driven by switch signals from the AWG. The NV center is resonantly excited, and fluorescence collection is enhanced by the SIL. This allows high-fidelity single-shot readout~\cite{Robledo2011Highfidelity}, achieving a fidelity of $(87.53\pm0.74)\%$ for both the bright state $\ket{0}_\text{e}$ and the dark state $\ket{1}_\text{e}$ (the crossing point in Fig.~\ref{fig:readoutFidelity}\textbf{a}). In each measurement, the electron spin is assigned to $\ket{0}_\text{e}$ if at least one photon is detected within the detection window; otherwise, it is assigned to $\ket{1}_\text{e}$.

\subsection{Energy level of the electron spin}

\begin{figure}[t]
\centering
\includegraphics[width=0.6\linewidth]{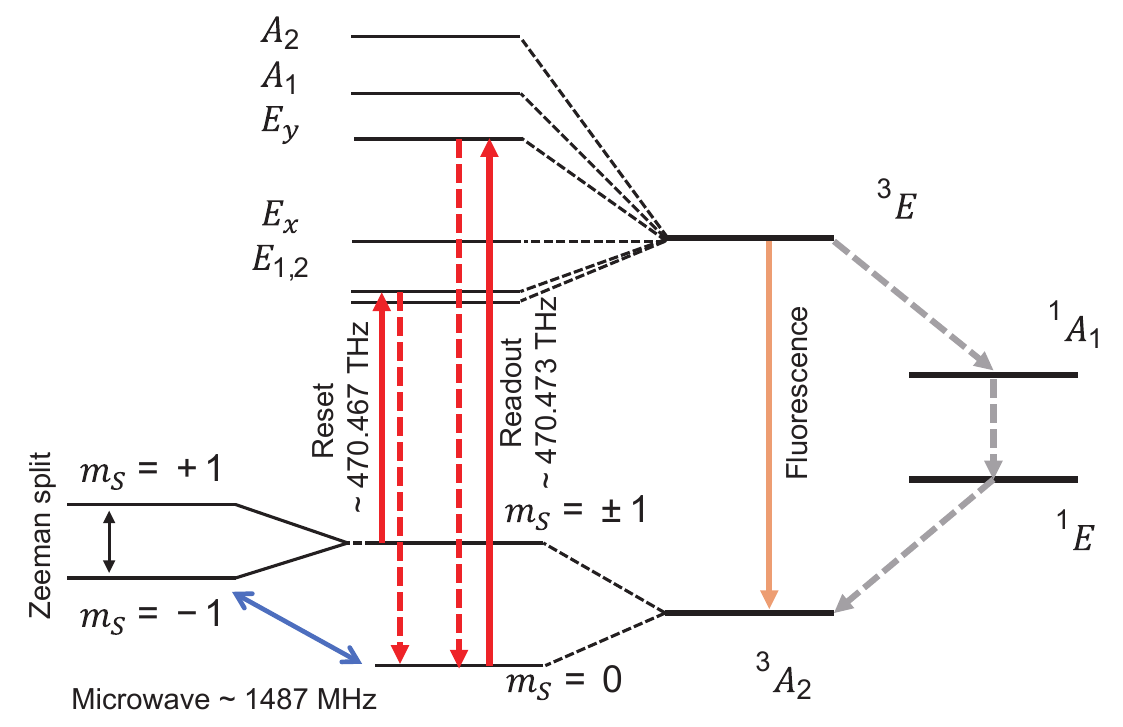}
\caption{
\textbf{Energy-level diagram of the NV center electron spin.}
The ground-state manifold consists of three spin sublevels, denoted $\ket{m_S=0}$ and $\ket{m_S=\pm 1}$. In the absence of an external field, the $\ket{m_S=\pm 1}$ states are separated from $\ket{m_S=0}$ by the zero-field splitting. In our setup, an external magnetic field of approximately 495~G lifts the degeneracy of the $\ket{m_S=\pm 1}$ states via the Zeeman effect. Resonant microwave fields drive transitions within this ground-state manifold.
The NV center can be pumped to the excited states with a lifetime of approximately 11~ns via resonant laser near 637~nm. We drive an approximate cycling transition $\ket{m_S=0}\longleftrightarrow \ket{E_y}$ for state readout. Reset of the NV center to the $\ket{m_S=0}$ ground state is accomplished by exciting the $\ket{m_S=\pm 1} \longleftrightarrow \ket{E_{1,2}}$ transitions, from which the system can decay partially to $\ket{m_S=0}$ via the metastable $^1A_1$ state. 
}
\label{fig:energyLevel}
\end{figure}

The energy-level structure of the negatively charged NV center~\cite{Lenef1996Electronic,Doherty2012Theory} is shown in Fig.~\ref{fig:energyLevel}. Before each experiment, the NV center is initialized in the negatively charged state using a 532-nm laser. The charge state is then verified by exciting the spin states using two resonant 637-nm lasers while detecting the fluorescence. If the fluorescence count exceeded a predetermined threshold, the experiment proceeded. Otherwise, the NV center would be reinitialized by the green laser, followed by another verification. This FPGA device determines the charge state according to the detected fluorescence and feeds back to the AWG in real time (Fig.~\ref{fig:system}).

\subsection{Control of qubits}

The experiment involved ten solid-state spin qubits, one electron spin and nine nuclear spins associated with a single NV center.
The electron spin is manipulated by microwave pulses, while nuclear spins are controlled by means of dynamical decoupling techniques, which indirectly manipulate the nuclear spin through tailored microwave pulses applied to the electron spin~\cite{Taminiau2014Universal,Hou2019Experimental}.
Dynamical-decoupling-based nuclear spin gates have two essential parameters, $\pi$-pulse number $N$ and interpulse decay $\tau$. The gate parameters for the nine nuclear spins are listed in Tab.~\ref{tab:nucParams}, where $\tau_{\text{CX}}$, $N_{\text{CX}}$ denote the parameters for the controlled-$X\left(\pm\frac{\pi}{2}\right)$ gate, and $\tau_{\text{Z}}$, $N_{\text{Z}}$ denote those for the $Z\left(\frac{\pi}{2}\right)$ gate.

\begin{table}[htbp]
\centering
\caption{
\textbf{Parameters of nuclear spin gate based on dynamical decoupling.} The  interpulse decay ($\tau_{\text{CX}}$, $\tau_{\text{Z}}$) and $\pi$-pulse numbers ($N_{\text{CX}}$, $N_{\text{Z}}$) for the controlled-$X\left(\pm\frac{\pi}{2}\right)$ gate and single-bit rotation $Z(\frac{\pi}{2})$ are listed for all nuclear spins.
}
\label{tab:nucParams}
\begin{tabular}{ccccc}
\toprule
nuclear index & $\tau_{\text{CX}}$ (ns) & $N_{\text{CX}}$ & $\tau_{\text{Z}}$ (ns) & $N_{\text{Z}}$\\
\midrule
2 & 7637 & 38 & 50 & 4\\
3 & 20460 & 14 & 37 & 4\\
4 & 5484 & 28 & 30 & 4\\
5 & 30435 & 26 & 50 & 4\\
6 & 7400 & 56 & 44 & 4\\
7 & 28426 & 56 & 38 & 4\\
8 & 29807 & 34 & 26 & 4\\
9 & 11471 & 56 & 46 & 4\\
10 & 18621 & 64 & 46 & 4\\
\bottomrule
\end{tabular}
\end{table}

\section{Experimental data collection and analysis}

\subsection{Data collection for training machine learning models}

\begin{figure}[t]
\centering
\includegraphics[width=0.7\textwidth]{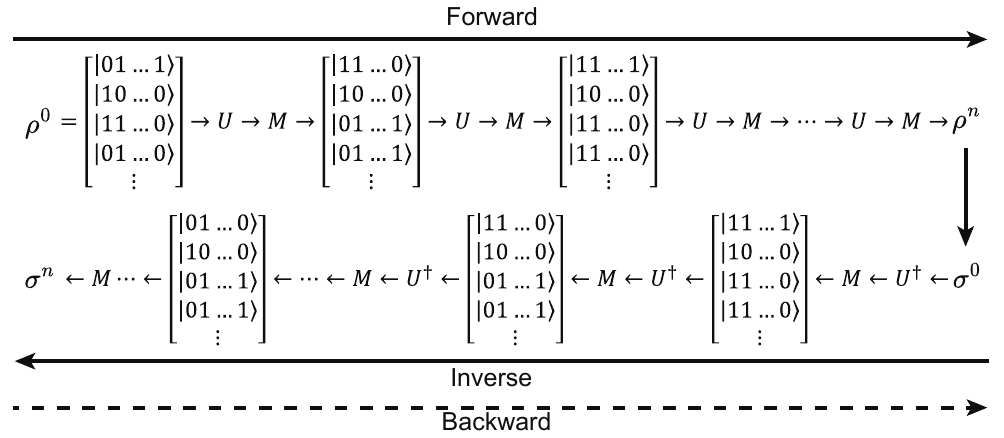}
\caption{
\textbf{Workflow of the thermodynamic protocol.}
The scheme start with an initial thermal state $\rho^0$ represented by a state ensemble (states in the brackets).  The system evolves under the unitary $U = \prod_{i=N}^1\mathcal{U}_i$ and subsequently undergoes the projective measurement $M$ of all qubits in the Pauli-$Z$ basis, resulting in another state ensemble. This process is repeated multiple cycles, yielding the forward evolution trajectories. 
The backward evolution begins with the final state of the forward process, $\rho^n = \sigma^0$, and proceeds under the time-reversed unitary $U^\dagger$.
}
\label{fig:experSeq}
\end{figure}

The process for acquiring evolution trajectories is illustrated in Fig.~\ref{fig:experSeq}. Each experimental trajectory begins with a thermal state of ten qubits, each assigned with a specific inverse temperature: $\beta_\text{e} = 1$ for the electron spin and $\beta_i = -1$ for nuclear spins 1 to 9. The system then evolves under the unitary operator $\prod_{i=N}^1\mathcal{U}_i$, with $\mathcal{U}_i = \exp\left[-\mathrm{i} \left( X_\text{e} X_i + Y_\text{e} Y_i\right)\Delta t\right]$, followed by projective measurements of all qubits in their Pauli-$Z$ basis. Repeating the unitary evolution and measurements multiple cycles and recording the sequential outcomes yields forward trajectories $\boldsymbol{\rho}^F=\{\rho^k\}_{k=0}^n$. Backward trajectories $\overline{\boldsymbol{\sigma}^R}=\{\sigma^{n-k}\}_{k=0}^n$ is obtained similarly by replacing $\prod_{i=N}^1\mathcal{U}_i$ with $\prod_{i=1}^N\mathcal{U}_i^\dagger$. The initial state of each backward trajectory is the final state of its corresponding forward trajectory. To directly compare the forward and backward processes, we reverse the temporal order of the recorded backward outcomes to obtain trajectories representing time-reversed evolution.

Ideally, recording full trajectories requires quantum non-demolition measurements, which are not feasible in our experimental setup due to the destructive nature of our measurements. To circumvent this issue, we adopt a state re-preparation approach, in which spin qubits are re-initialized according to the recorded outcomes after each measurement. This approach enables a simplified data-taking procedure. Concretely, we first prepare initial quantum states via probabilistic sampling, followed by unitary evolution under $\prod_{i=N}^1\mathcal{U}_i$. We then perform projective measurements on all qubits, record the outcomes, and re-initialize each qubit based on its individual measurement outcome to facilitate subsequent experimental operations. Each evolution step performed on the quantum processor only involves state initialization, unitary evolution, and readout. 

We simplify the data-taking procedure by performing single-step evolution runs with various initial states. For each run, we record the initial state and the measurement outcomes, compiling a large dataset. A full evolution trajectory is then constructed by randomly sampling single-step runs from this dataset, ensuring that the measurement outcome of one step matches the initial state of the subsequent step. 
This entire workflow is fully equivalent to the protocol described above.

The step-by-step construction of the forward dataset is elaborated below. The backward dataset is constructed using an identical procedure. The raw dataset is a matrix with dimensions $[650{,}000 \times 2]$. In this matrix, the first column contains initial quantum states (ranging from $\ket{00\dots 0}$ to $\ket{11\dots 1}$). The second column contains the corresponding measurement outcomes after unitary evolution ($\prod_{i=N}^1\mathcal{U}_i$) and projective measurement. The size of $650{,}000$ is estimated based on the number of ``initialize all spins $\rightarrow U \rightarrow Z$-basis measurements on all spins'' units required to assemble $45{,}000$ forward trajectories. The distribution of initial states in the raw dataset is predetermined via numerical simulation. This enables us to estimate both the number and distribution of units needed to generate the desired number of trajectories prior to experiments. Due to the randomness of measurements, the number of generated trajectories may be fewer than expected. This requires repeated experiments to obtain a sufficiently large dataset. To construct the forward dataset, $45{,}000$ data points are sampled from the raw dataset to form trajectories from step 1 to step 2. Using the quantum states at step 2 as new starting points, another $45{,}000$ data points are extracted to construct trajectories from step 2 to step 3. This iterative process is repeated to form the complete forward dataset. Importantly, each entry in the raw dataset can only be selected once (i.e., no reuse of data points).

Without intermediate measurements, both the forward and backward evolutions are governed by the same unitary $U$, making the two processes indistinguishable. The introduction of projective measurements breaks this time-reversal symmetry, thereby rendering the forward and backward trajectories distinguishable.

\begin{figure}[t]
\centering
\includegraphics[width=1.0\textwidth]{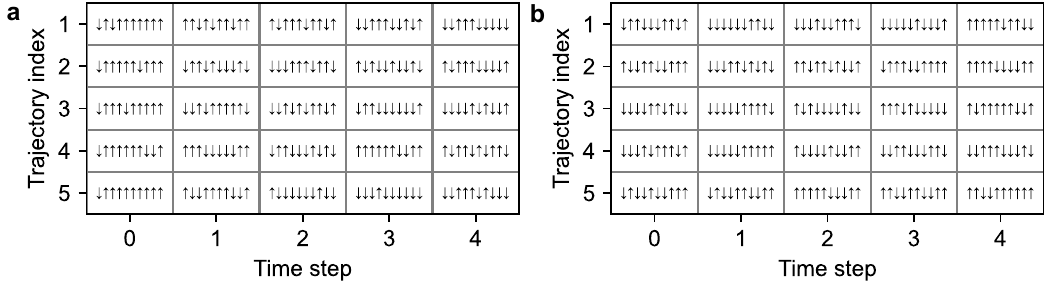}
\caption{
\textbf{Examples of experimental forward and backward trajectories.}
Five forward trajectories (\textbf{a}) and five backward trajectories (\textbf{b}) randomly selected from the experimental dataset. Each trajectory consists of the initial state and the final states of step 1 to step 4. Spin states are marked by arrows where $\uparrow$ corresponds to $\ket{0}$ and $\downarrow$ corresponds to $\ket{1}$.
}
\label{fig:exp_data_5}
\end{figure}

Using machine learning, we classify forward and backward state evolution trajectories, demonstrating that this approach can extract key features of physical processes---such as entropy increase, energy flow, and the irreversibility introduced by quantum measurement---from stochastic trajectory data. To illustrate this capability, we randomly selected five forward and five backward trajectories from the full dataset of $90{,}000$ experimental trajectories. 
For better visualization, one representative forward trajectory and one backward trajectory are shown in Fig.~\ref{fig:exp_and_gen_data_1}\textbf{a} and \textbf{b}, respectively. One may find it is hard to distinguish forward and backward trajectories by seeing the patterns in these example trajectories in Fig.~\ref{fig:exp_data_5}. 
However, one could perform a thermodynamic analysis to discriminate these trajectories. This indicates that the machine-learning model is able to extract fundamental physical principles from the dataset of $90{,}000$ trajectories.

\begin{figure}[t]
\centering
\includegraphics[width=0.8\textwidth]{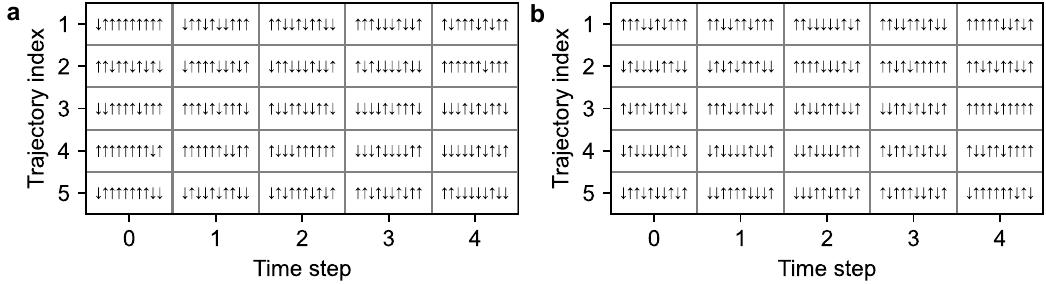}
\caption{
\textbf{Examples of generated trajectories by neural network.}
Five forward trajectories (\textbf{a}) and five backward trajectories (\textbf{b}) randomly selected from the generated dataset. Each trajectory consists of the initial state and the final states of step 1 to step 4. Spin states are marked by arrows where $\uparrow$ corresponds to $\ket{0}$ and $\downarrow$ corresponds to $\ket{1}$.
}
\label{fig:gen_data_5}
\end{figure}

\begin{figure}[t]
\centering
\includegraphics[width=0.8\textwidth]{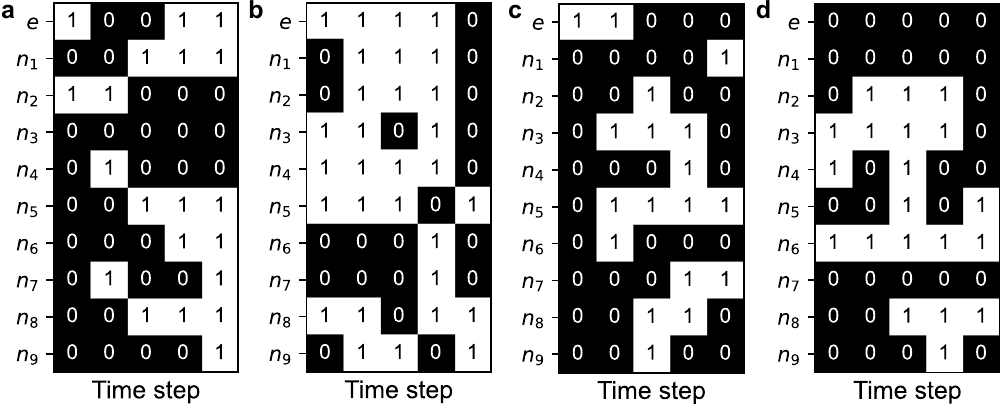}
\caption{
\textbf{Examples of experimental and neural-network generated trajectories.}
\textbf{a}, To more clearly illustrate the quantum state evolution process, we select the first forward experimental trajectory in Fig.~\ref{fig:exp_data_5}\textbf{a}. The horizontal axis denotes time step, and the vertical-axis denotes distinct qubits. $e$ denotes the electron spin, and $n_i$ denotes the $i$-th nuclear spin.
\textbf{b}, The first experimental backward trajectory in Fig.~\ref{fig:exp_data_5}\textbf{b}.
\textbf{c}, The first generated forward trajectory in Fig.~\ref{fig:gen_data_5}\textbf{a}.
\textbf{d}, The first generated backward trajectory in Fig.~\ref{fig:gen_data_5}\textbf{b}.
}
\label{fig:exp_and_gen_data_1}
\end{figure}

\subsection{Generative model}

To demonstrate that neural networks is able to capture the physical laws underlying the arrow of time, we trained a diffusion-based neural network to generate trajectories. As shown in Fig.~4 of the main text, the energy and entropy calculated from the generated data match the original experimental data, indicating that the network has effectively learned the key features of the arrow of time.
We select five forward trajectories and five backward trajectories from the generated dataset for visualizing the evolution of each qubit, as depicted in Fig.~\ref{fig:gen_data_5} and Fig.~\ref{fig:exp_and_gen_data_1}\textbf{c}, \textbf{d}.

\subsection{Unitary characterization}

\begin{figure}[t]
\centering
\includegraphics[width=0.75\textwidth]{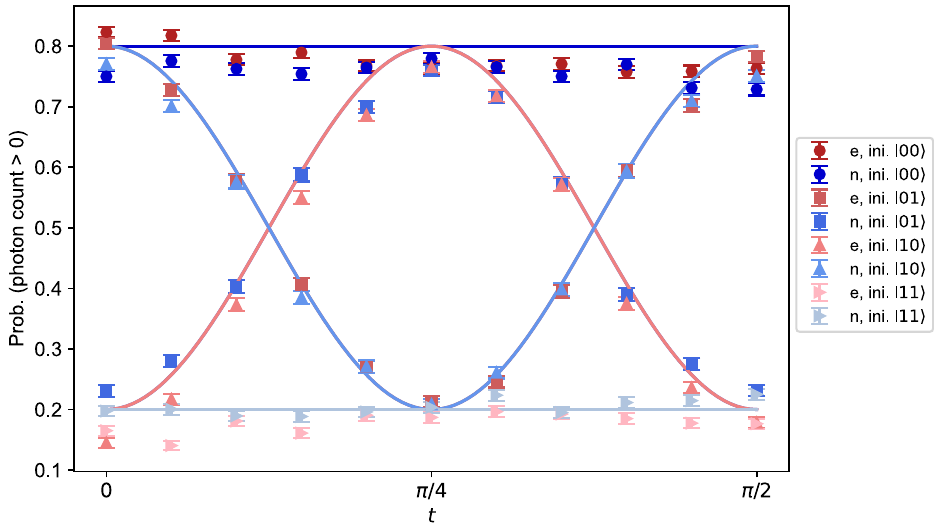}
\caption{
\textbf{Characterization of unitary $\mathcal{U}_i(t)=\exp\left[-\mathrm{i}\left(X_\text{e}X_i+Y_\text{e}Y_i\right)t\right]$}.
Plots of single spin population of the $i$th nuclear spin (blue symbols) the electron spin (red symbols) as a function of evolution time $t$. Four distinct initial two-spin states (see legend) are considered. The solid lines are numerical simulations of quantum state evolution governed by the unitary operator $\mathcal{U}_i(t)$, which are in good agreement with the experimental data.
The circuit implementing $\mathcal{U}_i(t)$ is shown in Fig.~2\textbf{c} of the main text, except that $\Delta t$ is replaced by $t$.}
\label{fig:tomo}
\end{figure}

The unitary operator in the experimental sequence is defined as $\mathcal{U}_i = \exp\left[-\mathrm{i} \left( X_\text{e} X_i + Y_\text{e} Y_i\right)\Delta t\right]$. There is no direct interaction between different nuclear spins, i.e., each nuclear spin interacts only with the electron spin.
To verify the correctness of the experimental sequence $\mathcal{U}_i$, we prepare four different initial states: $\ket{0}_\text{e}\ket{0}_i$, $\ket{0}_\text{e}\ket{1}_i$, $\ket{1}_\text{e}\ket{0}_i$, and $\ket{1}_\text{e}\ket{1}_i$. Each state then undergoes the designed sequence. Fig.~\ref{fig:tomo} shows the results of numerical simulations and the corresponding experimental data. The experimental results agree with theoretical expectations. Experimental infidelities arise primarily from imperfections in quantum gates and single-shot readout (Fig.~\ref{fig:readoutFidelity}\textbf{a}).

\end{document}